\documentclass[journal]{IEEEtran}

\usepackage{cite}
\usepackage{amsmath}
\usepackage{graphicx}
\usepackage{subcaption}
\usepackage{multirow}
\usepackage{comment}

\begin{document}

\title{Empirical Analysis of Near-Field Beam Shaping for Blockage Events}

\author{Jy-Chin~Liao,
        Volodymyr~Vizitiv,
        and~Edward~W.~Knightly,~\IEEEmembership{Fellow,~IEEE}%
\thanks{Jy-Chin Liao, Volodymyr Vizitiv, and Edward W. Knightly are with the Department of Electrical and Computer Engineering, Rice University, Houston, TX 77005, USA (e-mail: ll105@rice.edu; vv38@rice.edu; knightly@rice.edu).}}%


\maketitle

\begin{abstract}
Millimeter-wave (mmWave) to sub-terahertz (sub-THz) links can realize high data rates, yet are highly vulnerable to dynamic blockages due to high directivity, limited multipath, and dependence on line-of-sight (LoS) propagation. While structured and self-healing beams are promising solutions, it remains unclear how beam types perform under dynamic blockage. In this paper, we present an empirical analysis of structured near-field beams using two transmission policies. At one extreme, we study beam resilience without obstacle adaptation. At the other extreme, we empirically optimize beam parameters to study the limits of multiple beam types under idealized adaptation. We employ numerical analysis based on the angular spectrum method (ASM) and experimental validation using a sub-THz time-domain spectroscopy (TDS) platform to characterize beam performance over complete blockage events. For the scenarios considered, despite lacking self-healing, focused beams provide the strongest resilience when beam parameters are not adapted during a blockage event (i.e., when all beams are optimized only \emph{a priori}).  In contrast, when optimally adapted to obstacle position, curved beams provide the best performance. Lastly, although Bessel beams benefit from self-healing, this property alone can be insufficient to outperform optimized curved and focused beams. These findings provide guidance for blockage-aware beam selection and adaptation protocols.

\end{abstract}

\begin{IEEEkeywords}
Near-field communication, beam shaping, dynamic blockage, structured beams, metasurfaces.
\end{IEEEkeywords}

\section{Introduction}
Millimeter-wave (mmWave) to sub-terahertz (sub-THz) communication is a promising candidate for future wireless networks due to high bandwidth for ultra-high-capacity links~\cite{akyildiz2014terahertz, koenig2013wireless}. Unfortunately, short wavelengths and limited multipath propagation make such links highly vulnerable to obstruction by indoor objects such as human bodies, furniture, and moving obstacles~\cite{shurakov2023empirical, shurakov2023dynamic, doeker2025human}. Blockage degradation depends not only on line-of-sight (LoS) obstruction, but also on obstacle size, location, material response, diffraction, and the spatial structure of the transmitted beam. This beam dependence is especially important in the radiative near-field region, where electrically large apertures can synthesize non-planar wavefronts and spatially structured beams~\cite{petrov2023mobile, singh2023wavefront, zhang20236g, nepa2017near}. Unlike far-field steering, which mainly controls propagation direction, near-field beam shaping can redistribute energy over both range and angle, producing distinct main-lobe, side-lobe, and trajectory characteristics. As a result, different near-field beam types interact with obstacles through different combinations of shadowing, diffraction, and residual energy coupling~\cite{nepa2017near, liao2025terafocus, durnin1988comparison, reddy2023ultrabroadband, bodet2024sub, guerboukha2024curving}.

In this paper, we present an empirical analysis of near-field beam shaping under dynamic blockage, examining how different beam structures and beam adaptation strategies affect link performance. Unlike prior work that focuses on a single beam type, a static blockage geometry, or a specific beam adaptation algorithm, our goals are to understand the fundamental blockage resilience of structured near-field beams and to study an empirical upper bound on the performance gain achievable through beam adaptation. To this end, we define a \textit{blockage event} as the duration between when a blockage enters and exits the transmitter-receiver propagation path. We then evaluate performance over blockage \emph{events} rather than for static blockage \emph{states}.

We consider focused, Bessel, and curved beams as exemplary near-field beams, as well as (far-field optimized) steered beams as a baseline. For each beam type, parameter selection is critical. For example, a curved beam may have to switch from curving left to curving right during a blockage event. Moreover, while it may seem that a focused beam need not adapt to an obstacle when the receiver does not move and hence the focal point is unchanged, we will show that changing both focal length and steering angle can improve obstacle resilience.  Therefore, we represent each beam type using a beam-specific parameter vector and discretize its parameter space. To optimize beam parameters, we perform an exhaustive search over this parameter space. 

Based on this framework, we define two transmission policies to evaluate each beam type. Under the first policy, the \textit{adaptation-free policy}, each beam type uses the parameters that maximize its \emph{blockage-free} signal-to-noise ratio (SNR), and this configuration remains fixed throughout the blockage event. This policy quantifies the inherent blockage resilience of each beam type and shows whether a configuration optimized without blockage remains effective when the propagation path becomes blocked. In contrast, with the second \textit{adaptive upper-bound policy}, beam parameters are empirically re-optimized at each blockage position. By nearly continuously re-configuring parameters to always maximize SNR, this policy defines an empirical upper bound on what an adaptive beam can achieve, as real-world beam adaptation will incur feedback delays, overhead, imperfect channel state estimates, etc. 


Next, we use the two policies and the proposed framework to conduct a numerical analysis to investigate how beam type and adaptation policy affect blockage resilience for various blockage conditions. Due to our use of exhaustive search over beam parameters and types, we employ the angular spectrum method (ASM) as an efficient computational method. We first analyze the instantaneous SNR throughout a blockage event to study how different beam structures interact with a moving blockage and how beam adaptation improves the received performance. We then introduce event-level metrics to compare beam performance under different blockage sizes and locations. Additionally, we record how each beam type adapts its control parameters under the adaptive upper-bound policy. This helps explain when and how certain beam types perform better under specific blockage conditions. 
We find that blockage resilience depends on a complex combination of the beam's field-propagation properties, its ability and need to adapt, and the blockage conditions. For instance, focused beams provide the strongest resilience when beam parameters remain fixed, achieving up to 10 dB higher average SNR than the other beam types, while curved beams achieve the largest performance recovery through trajectory adaptation, with an adaptation gain of up to 13 dB. Moreover, the ``best'' beam type changes with blockage conditions. For example, curved beams perform best under small blockages where trajectory avoidance is effective, while focused beams are superior when blockage size increases and diffraction-assisted receiver coupling becomes more important. Surprisingly, although Bessel beams benefit from the self-healing property, even when optimized, they do not outperform focused or curved beams on average and remain up to 4 dB below the best-performing beam type. This result indicates that self-healing alone is insufficient to guarantee the strongest performance. Overall, the numerical results provide guidance on how to select the most suitable beam type for different blockage conditions while considering the benefits and overhead of beam adaptation.

Finally, we experimentally utilize the proposed framework using a sub-THz platform based on time-domain spectroscopy (TDS) and passive metasurface beam shapers. Since reconfigurable near-field beamformers (e.g., programmable metasurfaces and phased arrays) are not yet widely available, we realize different beam types and beam parameters using multiple static metasurface designs. 
The measurements confirm the numerical trends. Specifically, focused beams exhibit the strongest resilience without adaptation, whereas non-adaptive curved beams experience larger degradation during the part of the blockage event not favored by their curvature. On the other hand, curved beams achieve the greatest recovery through trajectory adaptation. Likewise, although Bessel beams benefit from their self-healing field structure, the measurements also demonstrate that self-healing does not produce the best blocked-link performance. The agreement between numerical analysis and experiments validates that the proposed empirical framework captures the dominant beam-dependent blockage behavior observed in practice. More importantly, the combined numerical and experimental results provide practical guidelines for balancing the benefits of adaptive beam control against protocol constraints such as overhead associated with blockage sensing, channel estimation, feedback, and hardware reconfiguration.

The remainder of this paper is organized as follows. Section~II reviews related work. Section~III presents the proposed empirical analysis framework. Section~IV provides the numerical evaluation; Section~V presents the experimental evaluation, and Section~VI concludes the paper.

\section{Prior Work}
\textbf{Dynamic Blockage in Sub-THz Links.}
Dynamic blockage has been extensively studied in mmWave and sub-THz networks~\cite{maccartney2017rapid}.
Prior experiments have characterized blockage loss, reflection-aided propagation, and human-motion-induced variations~\cite{shurakov2023empirical, shurakov2023dynamic, doeker2025human}. 
Moreover, blockage-aware near-field beamforming can be highly sensitive to outdated or inaccurate blockage information, since a mismatched beam trajectory may cause severe performance loss \cite{wang2026blockage}. These results highlight the importance of understanding the full blockage trajectory rather than only a single blockage state. However, existing dynamic-blockage studies do not study multiple near-field beam types interacting with a moving obstacle.

\textbf{Blockage Mitigation with Structured Beams.}
Bessel beams have been experimentally shown to reduce degradation under partial static blockage compared with conventional Gaussian beams by reconstructing their central lobe after obstruction~\cite{reddy2023ultrabroadband}. Curved terahertz (THz) beams have been experimentally shown to effectively route energy around obstacles by using geometry-specific beam designs~\cite{guerboukha2024curving}. Additionally, the use of curved Airy beams has been investigated to mitigate blockages in near-field THz communication links~\cite{islam2025blockage}. Furthermore, strategies for avoiding blockages with Airy beams have been explored through physics-informed learning, which optimizes beam trajectories based on various transmitter, receiver, and obstacle geometries~\cite{chen2025physics}. These studies demonstrate that structured beams can improve blockage resilience. 
Nevertheless, the studies in~\cite{reddy2023ultrabroadband,guerboukha2024curving} do not optimize the demonstrated beam parameters, making it unclear what optimal performance the considered beam types can achieve. They also do not investigate whether other structured beam types could outperform the selected beams under the same blockage conditions. Furthermore, all of the studies evaluate beam performance at a single blockage state rather than over a complete blockage event. As a result, they fail to determine when a blockage-avoidance beam should be preferred over other beam types or how its average performance evolves as the blocker crosses the propagation path.


\textbf{Blockage-Aware Near-Field Beam Design.}
A separate line of work studies beam design and parameter selection for obstructed near-field links. Recent analytical work has characterized Gaussian, focused, Bessel, and curving beams, providing design foundations for structured near-field beamforming~\cite{uchimura2025optimization}. Blockage-aware wavefront design has also been studied by deriving optimal transmit wavefronts under path blockage~\cite{uchimura2026optimal}, using learning-based Airy beam training from received beam patterns~\cite{weng2025learning}, and comparing focused, curved, Airy, and codebook-based beamforming strategies under obstructed near-field channels~\cite{zhang2026breaking}. However, these works primarily consider static blockage states, focus on certain blockage conditions, or rely on simulation-based evaluation, leaving open how multiple beam types behave over a complete dynamic blockage event.


\section{Analysis Framework for Blockage Events}

In this section, we develop a framework to evaluate near-field beam shaping under dynamic blockage. We first define the blockage-event geometry and beam-specific parameters. We then introduce the received-field and link-quality models, followed by the adaptation-free and the adaptive upper-bound policies.


\subsection{System Geometry and Beam Control}

\begin{figure}[t]
    \centering
    \includegraphics[width=1\linewidth]{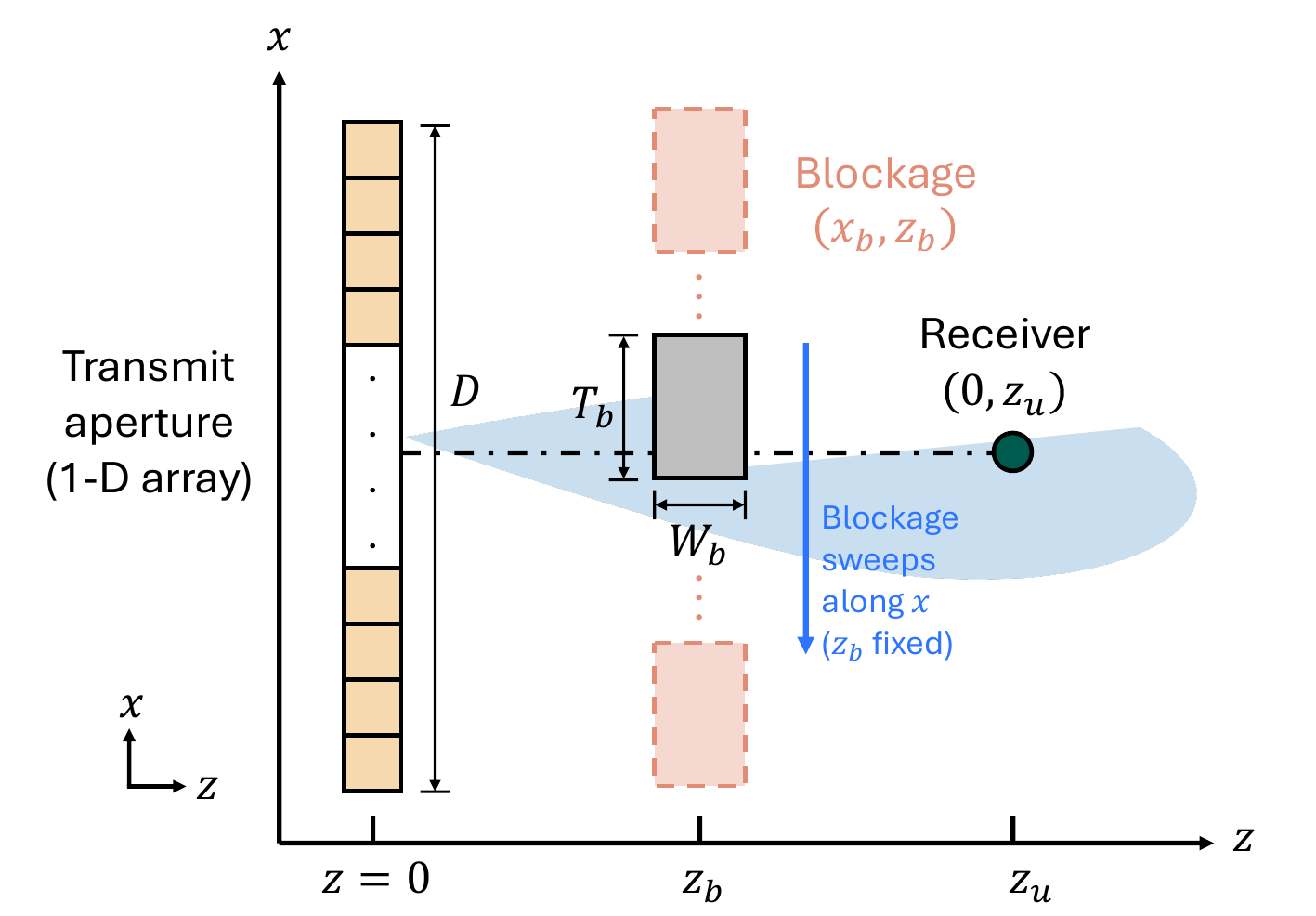}
    \caption{Illustration of the blockage-analysis geometry. The blocker center $x_b$ is swept laterally along the $x$-axis at a fixed longitudinal distance $z_b$.}
    \label{fig:blockage-analysis-setup}
\end{figure}

To ensure fair and consistent evaluation across various beam types, we first define the system geometry, blockage events, and beam control notation used throughout the analysis. Fig.~\ref{fig:blockage-analysis-setup} illustrates the analysis geometry in the $x$--$z$ plane. A one-dimensional transmit aperture of length $D$ is centered at $x=0$ in the plane $z=0$, with propagation along the $z$-axis. The receiver is located at broadside at $(x,z)=(0,z_u)$. A blockage with width $W_b$ and thickness $T_b$ is centered at $(x_b,z_b)$, where $z_b$ is its fixed longitudinal distance from the transmit aperture and $x_b$ is its lateral position. Each value of $x_b$ defines a blockage state $B(x_b)$, and sweeping $x_b$ laterally across the propagation region at fixed $z_b$ forms a \textit{blockage event}. 

\begin{figure*}[t]
\centering
\begin{subfigure}{0.24\textwidth}
    \centering
    \includegraphics[width=\linewidth]{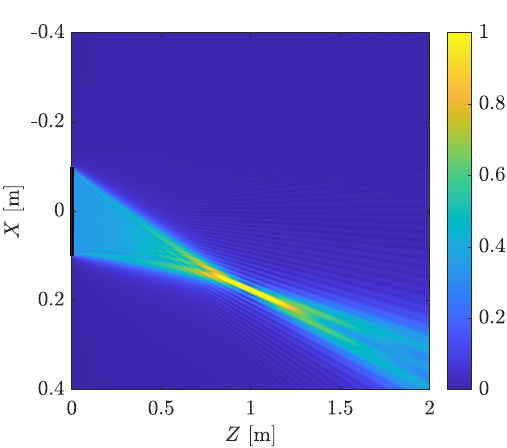}
    \caption{Focused beam}
    \label{fig:focused-beam}
\end{subfigure}
\hfill
\begin{subfigure}{0.24\textwidth}
    \centering
    \includegraphics[width=\linewidth]{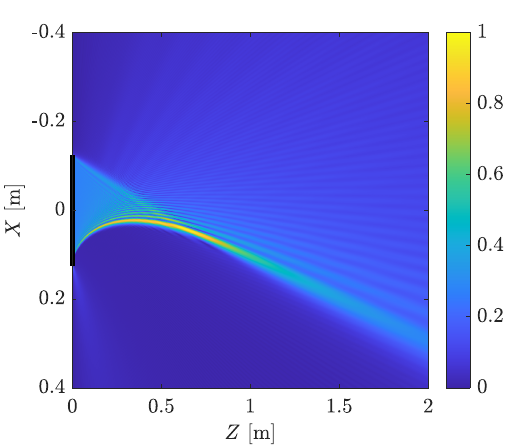}
    \caption{Curved beam}
    \label{fig:curved-beam}
\end{subfigure}
\hfill
\begin{subfigure}{0.24\textwidth}
    \centering
    \includegraphics[width=\linewidth]{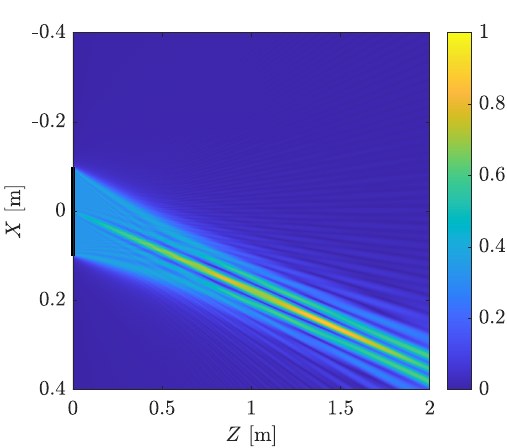}
    \caption{Bessel beam}
    \label{fig:bessel-beam}
\end{subfigure}
\hfill
\begin{subfigure}{0.24\textwidth}
    \centering
    \includegraphics[width=\linewidth]{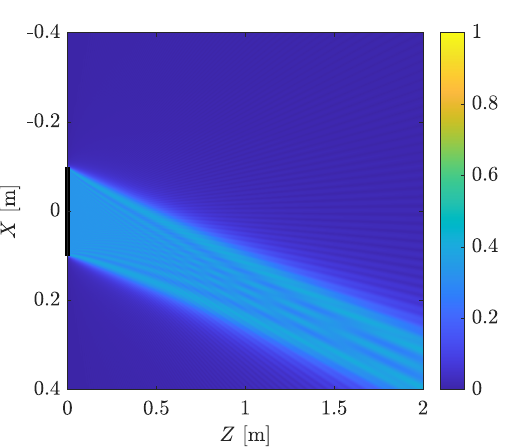}
    \caption{Steered beam}
    \label{fig:steered-beam}
\end{subfigure}

\caption{Normalized near-field electric-field (E-field) intensity distributions of the four candidate beam types.}
\label{fig:beam-shaping-toolbox}
\end{figure*}

We consider four representative beam types with distinct blockage-related characteristics: a focused beam with strong spatial energy concentration, a curved beam with a nonlinear propagation trajectory, a Bessel beam with a self-reconstructing field structure, and a conventional steered beam as the baseline. Their normalized near-field intensity distributions are shown in Fig.~\ref{fig:beam-shaping-toolbox}. The figure highlights localized focusing, curved propagation, the Bessel beam’s central core and surrounding side lobes, and conventional straight steering.

To describe how each beam type is controlled, we consider a phase-only, fixed-amplitude array model, following the near-field beamforming designs in prior works \cite{arora2022efficient,prado2022nearfield}. The position of the $n$-th antenna
is denoted by $x_n$. The wavenumber is $k=2\pi/\lambda$, where $\lambda$ is the wavelength, and the steering angle $\theta$ is measured from broadside. For beam type $s$, the phase applied to the $n$-th antenna is denoted by $\phi_{n,s}(\Theta_s)$, where $\Theta_s$ contains the corresponding beam-control parameters. We discretize the parameter space of each beam type into a finite set of candidate configurations, allowing all considered configurations to be empirically evaluated. We then select the best-performing configuration at each discretized blockage position $x_b$.

\textbf{Focused beams.}
The focused beam concentrates energy at a target location specified by angle and focal length \cite{nepa2017near,zhang20236g}. Such a range-dependent characteristic has been used to support high-rate mobile links \cite{liao2025terafocus} and user separation in the near-field \cite{wu2023multiple, hassan2026multi}. Its control parameter vector is $\Theta_{\mathrm{Focused}}=\{\theta,l\}$, where $\theta$ denotes the steering angle and $l$ the focal distance. The applied phase is
\begin{equation}
\phi_{n,\mathrm{Focused}}(\theta,l)
=
k\left(\sqrt{x_n^2+l^2}-l-x_n\sin\theta\right).
\end{equation}

\textbf{Curved beams.}
We consider curved beams that form a parabolic trajectory, enabling the dominant propagation path to bend around an obstacle \cite{guerboukha2024curving, chen2025physics,zhang2026breaking}. Their asymmetric spatial distribution and controllable trajectory offer an approach to blockage avoidance. We define their control parameter vector as
$\Theta_{\mathrm{Curved}}=\{\theta,l,\kappa\}$,
where $\theta$ controls the steering direction, $l$ the trajectory length, and $\kappa$ the trajectory curvature.

To realize a beam trajectory $g(z;\kappa)=\kappa\sqrt{z}$ for $0\le z\le l$, the applied phase is \cite{guerboukha2024curving}
\begin{equation}
\begin{aligned}
\phi&_{n,\mathrm{Curved}}(\theta,l,\kappa)
\\
&=-k\Bigg[
\int_{0}^{x_n}
\frac{g'(z_\xi;\kappa)}
{\sqrt{1+\left[g'(z_\xi;\kappa)\right]^2}}
d\xi 
+x_n\sin\theta
\Bigg],
\end{aligned}
\end{equation}
where $g'(z;\kappa)=\kappa/(2\sqrt{z})$ and $z_\xi$ is the point on the caustic whose tangent intersects the aperture at $\xi$. 

\textbf{Bessel beams.}
Bessel beams form a structured field with a central core and surrounding side lobes that support self-healing after partial obstruction \cite{durnin1988comparison,reddy2023ultrabroadband,bouchal1998self,aiello2014wave}. For steered Bessel beams, we define their control parameter vector as
$\Theta_{\mathrm{Bessel}}=\{\theta,\psi_b\}$,
where $\theta$ denotes the steering angle and $\psi_b$ controls the transverse energy spread. Following the phase-only design in \cite{reddy2023ultrabroadband}, the applied phase is
\begin{equation}
\phi_{n,\mathrm{Bessel}}(\theta,\psi_b)
=
k\left(|x_n|\sin\psi_b-x_n\sin\theta\right).
\end{equation}

\textbf{Steered beams.}
Steered beams serve as the conventional directional baseline as they are optimized for the far-field. We define their control parameter vector as
$\Theta_{\mathrm{Steered}}=\{\theta\}$,
with phase profile
\begin{equation}
\phi_{n,\mathrm{Steered}}(\theta)
=
-kx_n\sin\theta.
\end{equation}

\subsection{Received Field and Link-Quality Model}

Given a beam type $s$ with parameter vector $\Theta_s$, and a blockage position $x_b$,
we evaluate link quality under various blockage events by propagating the aperture function through the near-field region and computing the received power. We compute near-field propagation using ASM, which simplifies the computation of the Rayleigh-Sommerfeld integral by evaluating scalar diffraction in the Fourier domain \cite{born2013principles,goodman2017introduction,petrov2024wavefront,yazdnian2025nirvawave}. This enables efficient exhaustive-search parameter optimization for each blockage position.

For beam type $s$, the initial field across the transmit aperture is
\begin{equation}
E_s^{(0)}(x;\Theta_s)=a(x)e^{j\phi_s(x;\Theta_s)},
\end{equation}
where $a(x)$ is the amplitude profile (1 in our case) and $\phi_s(x;\Theta_s)$ is the corresponding phase profile. In the considered discrete array, the field is evaluated at the element positions $x_n$, with phase $\phi_{n,s}(\Theta_s)$. The amplitude profile is normalized such that all beam types and parameter configurations use the same transmit power.

The propagation axis is discretized into planes $z_m=m\Delta z$. Over each axial step $\Delta z$, ASM applies the transfer function
\begin{equation}
H(k_x,\Delta z)=\exp\left(-j\Delta z\sqrt{k^2-k_x^2}\right),
\end{equation}
where $k_x$ is the transverse spatial wavenumber.

The blockage state $B(x_b)$ is represented by a spatial attenuation mask $M_{B(x_b)}(x,z)$ over the corresponding blocker region $\Omega_{B(x_b)}$:
\begin{equation}
M_{B(x_b)}(x,z)=
\begin{cases}
\alpha_B, & (x,z)\in\Omega_{B(x_b)},\\
1, & \text{otherwise},
\end{cases}
\end{equation}
where $\alpha_B\in[0,1]$ represents the attenuation introduced by the blocker, with $\alpha_B=0$ corresponding to a fully opaque blocker.

The blocked field is then propagated iteratively as
\begin{align}
E_s^{(m+1)}&\bigl(x;\Theta_s,B(x_b)\bigr)=M_{B(x_b)}(x,z_{m+1})\times\\ \nonumber
&\mathcal{F}^{-1}\left\{H(k_x,\Delta z)\mathcal{F}\left[E_s^{(m)}\bigl(x;\Theta_s,B(x_b)\bigr)\right]\right\},
\end{align}
where $\mathcal{F}$ and $\mathcal{F}^{-1}$ denote the Fourier and inverse Fourier transforms along the $x$-dimension. The mask is applied at each propagation plane intersecting the finite axial thickness $T_b$ of the blocker.

At the receiver plane $z=z_u$, the received power is computed over the receiver aperture $\mathcal{L}_{\mathrm{rx}}$ as
\begin{equation}
P_{\mathrm{rx},s}\bigl(\Theta_s,B(x_b)\bigr)=\int_{\mathcal{L}_{\mathrm{rx}}}\left|E_s\bigl(x,z_u;\Theta_s,B(x_b)\bigr)\right|^2\,dx.
\end{equation}
Thus, we can compute the corresponding received SNR as
\begin{equation}
\gamma_s\bigl(\Theta_s,B(x_b)\bigr)=10\log_{10}\left(\frac{P_{\mathrm{rx},s}\bigl(\Theta_s,B(x_b)\bigr)}{P_n}\right),
\end{equation}
where $P_n$ is the receiver noise power. Because the receiver aperture, bandwidth, and noise power are fixed across all beam types, parameter configurations, and blockage states, maximizing the received power is equivalent to maximizing the received SNR.

\subsection{Transmission Policies}

The received model in the previous subsection allows us to evaluate the SNR of any beam type and parameter configuration under any blockage state. We now use this model to define two transmission policies for analyzing a dynamic blockage event. The goal is not to propose a specific real-time adaptation algorithm, but to separate two fundamental effects: how well a chosen beam survives blockage when it is kept fixed, and how much performance can be recovered if the beam parameters are optimally reconfigured.

This distinction is important because a beam type can perform well for different reasons. A beam may be robust if its clear-channel configuration continues to deliver energy to the receiver even when the blocker moves across the link. On the other hand, a beam may be vulnerable with its original configuration, but highly recoverable if its parameters are reconfigured to redirect, refocus, or redistribute energy around the blocker. To study these two effects, we define two policies: the \textit{adaptation-free policy} and \textit{adaptive upper-bound policy}. The former represents no real-time beam adaptation during the blockage event, while the latter represents ideal adaptation in which beam parameters are empirically re-optimized (via ASM) for each blockage position. 


In particular, each beam type is evaluated under the two policies over a discrete set of candidate parameter configurations. Let $\Omega_s$ denote the candidate set for beam type $s$. 
For example, the steering angle is discretized into a grid over an angular search range, while beam-specific grids control focal depth, curved trajectory, or transverse energy distribution. Each candidate set is defined to span the full parameter range over which changes in the beam configuration can affect power at the receiver. 

Let $\mathcal{X}$ denote the discrete set of blocker positions along the blockage trajectory. For a blocker position $x\in\mathcal{X}$, let $B(x)$ denote the corresponding blockage state, and let $B_0$ denote the unblocked clear-channel state.

\textbf{Adaptation-free policy.}
The adaptation-free policy models a transmitter that selects its beam configuration before the blockage event begins and keeps this configuration fixed as the blocker moves. Namely, the link is optimized for the clear channel but does not adapt to subsequent blockage dynamics. Thus, for each beam type $s$, the fixed configuration is selected via ASM-based exhaustive search over $\Omega_s$ in the unblocked state:
\begin{equation}
\Theta_s^{\mathrm{AF}}
=\arg\max_{\Theta_s\in\Omega_s}
\gamma_s(\Theta_s,B_0).
\end{equation}
Once selected, $\Theta_s^{\mathrm{AF}}$ is not updated during the blockage event. The instantaneous SNR under the adaptation-free policy can then be computed with ASM as
\begin{equation}
\gamma_s^{\mathrm{AF}}(x)
=\gamma_s(\Theta_s^{\mathrm{AF}},B(x)),
\qquad x\in\mathcal{X}
\end{equation}
while the corresponding clear-channel SNR is
\begin{equation}
\gamma_{s,0}
=\gamma_s(\Theta_s^{\mathrm{AF}},B_0).
\end{equation}
These SNRs will also be experimentally measured in Section V. This policy, therefore, characterizes the inherent blockage resilience of the nominal clear-channel beam configuration. A high value of $\gamma_s^{\mathrm{AF}}(x)$ during blockage indicates that the selected beam type and parameters remain effective without requiring real-time adaptation.

\textbf{Adaptive upper-bound policy.}
The adaptive upper-bound policy models a transmitter that nearly instantaneously adapts to the blockage. In particular, at each blockage state, the transmitter re-selects its parameters (i.e., reconfigures its aperture function or antenna weights) to the parameters that ASM indicates yield the highest instantaneous SNR. Thus, in contrast to the adaptation-free policy, which commits to a single clear-channel configuration, this policy repeats the search over $\Omega_s$ independently at every blockage position so that for each $x\in\mathcal{X}$, the selected parameter configuration is
\begin{equation}
\Theta_s^{\mathrm{UB}}(x)
=\arg\max_{\Theta_s\in\Omega_s}
\gamma_s(\Theta_s,B(x)).
\end{equation}
The resulting instantaneous SNR is
\begin{align}
    \gamma_s^{\mathrm{UB}}(x)
&=\gamma_s(\Theta_s^{\mathrm{UB}}(x),B(x))\\
&=\max_{\Theta_s\in\Omega_s}
\gamma_s(\Theta_s,B(x)).
\end{align}

While not realizable in practice as this policy assumes perfect knowledge of the blockage state (e.g., ignoring channel sensing, feedback, computation, and hardware reconfiguration overhead), it serves as an empirical upper bound over the evaluated parameter set.

Note that the gap between $\gamma_s^{\mathrm{UB}}(x)$ and $\gamma_s^{\mathrm{AF}}(x)$ quantifies the maximum instantaneous SNR recovery available from beam-parameter adaptation for beam type $s$. A small gap indicates that the clear-channel beam configuration is already robust to blockage, while a large gap indicates that the beam type has substantial adaptation potential, but requires parameter reconfiguration to realize it.

\section{Numerical Analysis}

In this section, we numerically evaluate the selected set of near-field beam types under representative dynamic blockage scenarios using the proposed transmission policies. The evaluation considers both instantaneous SNR over the blockage event and event-level performance metrics that quantify the impact of the full blockage event.

\subsection{Numerical Setup}
While full-wave electromagnetic solvers can provide accurate field modeling, they are computationally expensive for the repeated evaluations required in this study. 
Therefore, we use ASM as described in Section III.B to compute near-field propagation and diffraction during the blockage event \cite{goodman2017introduction,yazdnian2025nirvawave}.
The field calculations are performed at a carrier frequency of $150$ GHz. The transmit aperture size is set to $D=0.5$ m, and the receiver is located at $1$ m along the propagation direction. To emulate a dynamic blockage event, the blockage center is swept laterally across the propagation region with a step size of $0.02$ m. Each sampled blockage position $x_b$ defines one blockage state $B(x_b)$ along the event.

The continuous aperture of size $D = 0.5$~m is evaluated over a computational window size of $1.2$~m, and the longitudinal resolution for ASM propagation is defined by $2000$ points. The blocker sweep ranges from $x = -0.4 - W_b/2$ to $x = 0.4 - W_b/2$ with a step size of $0.02$~m, modeled with a finite axial thickness $T_b = 0.05$~m and an opaque attenuation factor $\alpha_B = 0$. At the receiver, the aperture size is $\mathcal{L}_{\mathrm{rx}} = 0.05$~m, and the noise power $P_n$ is fixed across all evaluations.

The beam-parameter search spaces are summarized in Table~\ref{tab:beam-parameter-grid}. For each beam type, we sweep its corresponding parameter grid and compute the resulting SNR over the full blockage trajectory. Under the adaptation-free policy, each beam uses the configuration that gives the highest SNR under the clear-channel condition, found through an empirical exhaustive search over its full discretized parameter set. Under the adaptive upper-bound policy, the same empirical search is repeatedly performed at each blockage state to select the configuration that yields the highest SNR under the current blockage condition.
For example, as shown in Table \ref{tab:beam-parameter-grid}, the curved beam's blockage event is evaluated over $181$ steering angles, $11$ focal lengths, and $11$ curvature values, yielding $21,901$ total field calculations at each blockage position. Therefore, the overall computational cost scales linearly with the number of blockage positions sampled throughout the event.

\begin{table}[t]
    \centering
    \begin{tabular}{cll}
        \hline
        Beam type & Parameter & Search space \\
        \hline
        Steered 
        & $\theta$ 
        & $[-90^\circ,90^\circ]$, step $1^\circ$ \\
        \hline
        \multirow{2}{*}{Focused} 
        & $\theta$ 
        & $[-90^\circ,90^\circ]$, step $1^\circ$ \\
        & $l$ 
        & $[0,2]$~m, step $0.2$~m \\
        \hline
        \multirow{2}{*}{Bessel} 
        & $\theta$ 
        & $[-90^\circ,90^\circ]$, step $1^\circ$ \\
        & $\psi_b$ 
        & $[0^\circ,20^\circ]$, step $1^\circ$ \\
        \hline
        \multirow{3}{*}{Curved} 
        & $\theta$ 
        & $[-90^\circ,90^\circ]$, step $1^\circ$ \\
        & $l$ 
        & $[0,2]$~m, step $0.2$~m \\
        & $\kappa$ 
        & $[0,1]$, step $0.1$ \\
        \hline
    \end{tabular}
    \caption{Beam-parameter search spaces used in the numerical analysis.}
    \label{tab:beam-parameter-grid}
\end{table}

\subsection{Instantaneous SNR Analysis}

To understand how each beam type responds under dynamic blockage, we first study the instantaneous SNR as a moving blockage enters, intersects, and exits the near-field propagation region. This analysis captures the SNR evolution under both transmission policies, revealing where the dominant fades occur and how adaptation changes each beam’s blockage response.

\begin{figure}[t]
    \centering

    \begin{subfigure}{1\linewidth}
        \centering
        \includegraphics[width=\linewidth]{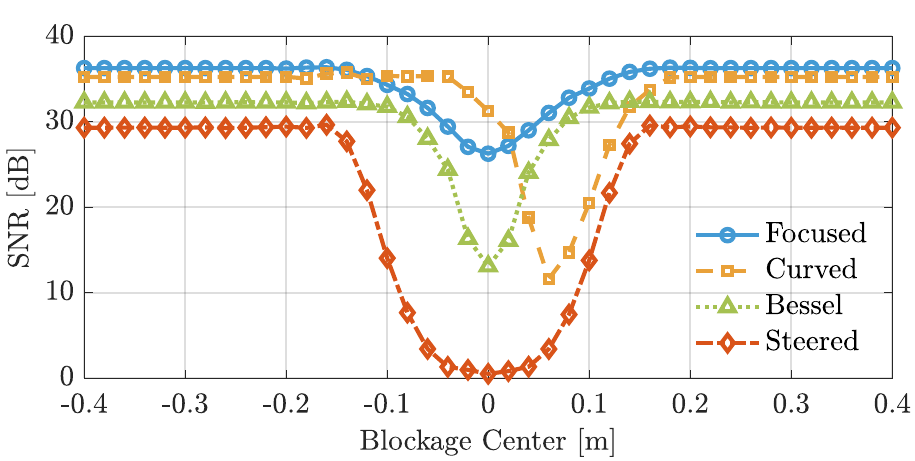}
        \caption{Adaptation-free policy}
        \label{fig:instantaneous-af}
    \end{subfigure}

    \vspace{0.5em}

    \begin{subfigure}{1\linewidth}
        \centering
        \includegraphics[width=\linewidth]{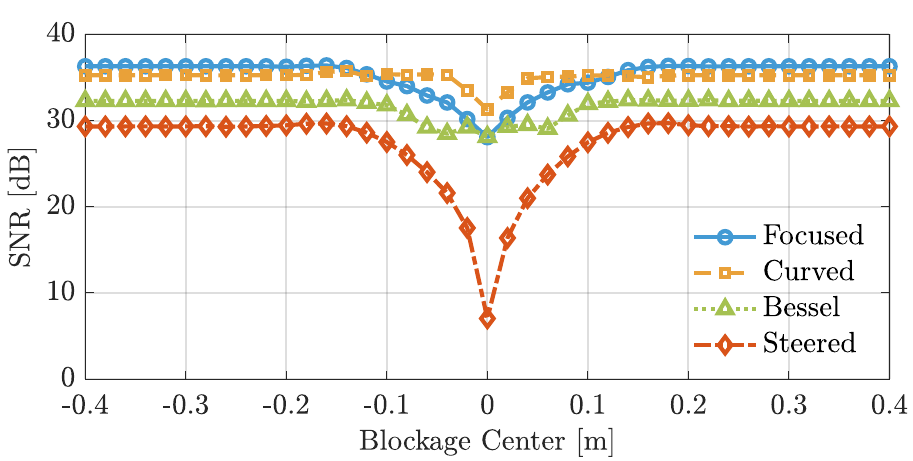}
        \caption{Adaptive upper-bound policy}
        \label{fig:instantaneous-ub}
    \end{subfigure}

    \caption{Instantaneous SNR during a lateral blockage sweep for the four candidate beam types.}
    \label{fig:instantaneous-rx-power}
\end{figure}

We consider a blockage scenario with blockage width $W_b=D/2=0.25$ m. The blockage is placed at $z_b=0.5$ m, midway between the transmitter and receiver, with the receiver located at $z_u=1$ m. To emulate a dynamic blockage event, the blockage is swept laterally across the propagation region, and the received SNR is computed along the blockage trajectory. Fig.~\ref{fig:instantaneous-rx-power} compares the resulting instantaneous SNR of the four candidate beam types under the adaptation-free and adaptive upper-bound policies.

Fig.~\ref{fig:instantaneous-af} shows the instantaneous SNR under the adaptation-free policy, where each beam uses the clear-channel configuration that maximizes the received SNR and keeps its parameters fixed throughout the blockage event. The SNR remains high when the blockage is outside the dominant propagation region, but decreases once the blockage enters the beam path. The depth, symmetry, and duration of the resulting fade vary substantially across beam types.

The baseline steered beam experiences both the worst clear-channel performance and the deepest SNR degradation when the blockage overlaps its main lobe. This is expected as the steered beam is optimized for the far field. However, perhaps surprisingly, the focused beam outperforms the Bessel beam despite only the Bessel beam having the self-healing property. Moreover, the focused beam has the best average performance over the event, indicating superior resilience without adaptation for this scenario. Yet, the curved beam outperforms the focused beam when the obstacle is on the left side of the aperture, at the center, or slightly to the right of center. This is because the clear-channel optimization selected a right-curving trajectory, thereby achieving the best performance by avoiding the obstacle when it is on the left. However, as the obstacle moves to the right (positive $x$), without adaptation, the curved beam simply curves the wrong way and performs poorly. 

\textit{Findings: Without adaptation, the focused beam provides the strongest overall resilience. While the Bessel beam benefits from self-healing, it cannot outperform the gains due to focusing in this scenario. Moreover, while the curving beam achieves the best performance for some obstacle positions, for other obstacle positions, a lack of adaptation to blockage position can degrade its performance to the worst of the three near-field beams studied. Finally, the far-field optimized steered beam performs the worst for all obstacle positions, even underperforming a curved beam that is curving the wrong way for the obstacle's position ($x > 0$). }

Fig.~\ref{fig:instantaneous-ub} depicts instantaneous SNR under the adaptive upper-bound policy, in which beam parameters are empirically re-optimized for each blockage position throughout the blockage event. Observe that compared with the adaptation-free policy, the deepest fades are reduced for all beam types. Moreover, the curved beam now has superior performance in the region near $x=0$ when the obstacle is near the center. This is because the curved beam can now adaptively bend either to the right or left and vary its curvature according to the obstacle position. For Bessel and focused beams, it may seem that there should be no adaptation gain as the focal length doesn't change as the obstacle moves and the Bessel beam is already designed to self-heal in the presence of an obstacle. However, the figure clearly indicates adaptation gains for both Bessel and focused beams: the Bessel beam achieves improvement by jointly adapting its steering and deflection angles, while the focused beam also benefits from adapting focal length and steering angle. 

\textit{Findings: All beam types benefit from adaptation to obstacle position \emph{despite} the receiver remaining stationary. In particular, all beams benefit from steering \emph{away} from the receiver in order to realize gains from diffraction off of the obstacle's edges. Moreover, the focused beam adapts its focal distance to enhance diffraction-assisted receiver coupling, whereas the curved beam continuously reshapes its propagation trajectory to bypass the moving blockage, achieving the largest adaptation gain. Although the Bessel beam jointly adapts its steering and deflection angles to exploit self-healing, it is still outperformed by the focused and curved beams in this blockage scenario, with the Bessel and focused beams achieving nearly identical performance only when the obstacle is centered. This indicates that self-healing alone is insufficient to guarantee the strongest blocked-link performance, and that effective blockage mitigation depends on how well each beam exploits its overall propagation characteristics.}

\subsection{Event-Level Metrics}
While instantaneous analysis explains how a beam responds at each blockage position, here we analyze SNR over the full blockage event. This facilitates the study of factors such as blockage size, material, and shape, as well as distance to the transmitter and receiver. To avoid bias in the beam-specific event-level averages, we define a beam-dependent \textit{blockage zone} based on the adaptation-free policy, so that each metric is computed only where the fixed clear-channel beam is meaningfully degraded.

Let $\mathcal{X}$ denote the set of blockage positions along the trajectory, $\gamma_s^{\mathrm{AF}}(x)$ denote the instantaneous SNR of beam type $s$ under the adaptation-free policy, and $\gamma_{s,0}$ denote its clear-channel SNR. We define the blockage zone as
\begin{equation}
    \mathcal{B}_{s,\tau}
    =
    \{x\in\mathcal{X}: \gamma_{s,0}-\gamma_s^{\mathrm{AF}}(x)\geq \tau\},
\end{equation}
where $\tau$ is the SNR-loss threshold. This definition identifies the beam-specific portion of the trajectory where blockage has a non-negligible impact.

Within $\mathcal{B}_{s,\tau}$, we evaluate two metrics. Note that these metrics apply only when $|\mathcal{B}_{s,\tau}| > 0$. When there is no blockage (e.g., $W_b/D = 0$), the blockage zone is empty, and we explicitly define the adaptive upper-bound SNR as the clear-channel SNR ($\overline{\gamma}_{s}^{UB} = \gamma_{s,0}$) and the adaptation gain as zero ($G_{s}^{adapt} = 0$). For $|\mathcal{B}_{s,\tau}| > 0$, the first metric is the \textit{adaptive upper-bound SNR}, defined as
\begin{equation}
    \overline{\gamma}_s^{\mathrm{UB}}
    =
    \frac{1}{|\mathcal{B}_{s,\tau}|}
    \sum_{x\in\mathcal{B}_{s,\tau}}
    \gamma_s^{\mathrm{UB}}(x),
\end{equation}
where $\gamma_s^{\mathrm{UB}}(x)$ is the instantaneous SNR under the adaptive upper-bound policy. This metric captures the best achievable event-level performance with ideal beam-parameter adaptation within the given beam-parameter spaces.

The second metric is the \textit{adaptation gain}, defined as
\begin{equation}
    G_s^{\mathrm{adapt}}
    =
    \frac{1}{|\mathcal{B}_{s,\tau}|}
    \sum_{x\in\mathcal{B}_{s,\tau}}
    \left(\gamma_s^{\mathrm{UB}}(x)-\gamma_s^{\mathrm{AF}}(x)\right).
\end{equation}
This metric quantifies the average SNR recovery provided by adaptation. Together, $\overline{\gamma}_s^{\mathrm{UB}}$ and $G_s^{\mathrm{adapt}}$ compare both achievable performance and adaptation benefit across blockage conditions.

\textbf{Effect of Blockage Size:} We first study how blockage size affects event-level performance, as larger obstacles increasingly challenge beam resilience, albeit differently for each beam type. Here, the blockage is placed at $0.5$~m, midway between the transmitter and the receiver, while the receiver is located at $1$~m. We vary the blockage width normalized by the transmit aperture size, $W_b/D$, where $D=0.5$~m, and consider $W_b/D\in\{0,1/8,1/4,1/2,1\}$. For each blockage width, event-level metrics are computed over the blockage zone using $\tau=0.1$~dB. When $W_b/D=0$, no blockage is present. Thus, $\overline{\gamma}_s^{\mathrm{UB}}$ corresponds to the clear-channel SNR and $G_s^{\mathrm{adapt}}=0$.

\begin{figure}[t]
\centering

\begin{subfigure}{0.48\linewidth}
    \centering
    \includegraphics[width=\linewidth]{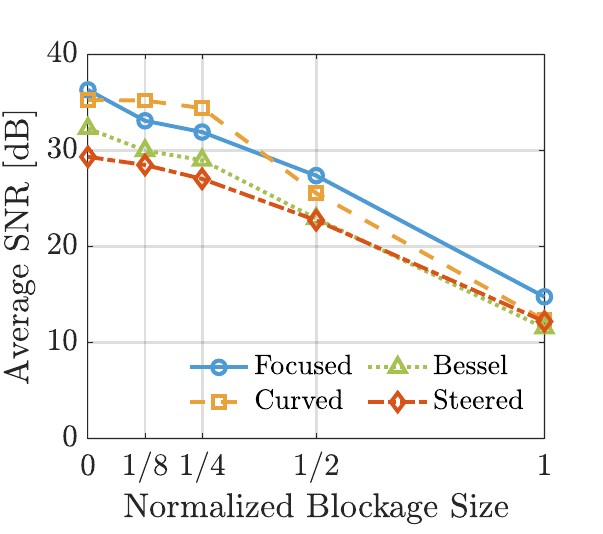}
    \caption{}
    \label{fig:avg-snr-blockage-size}
\end{subfigure}
\hfill
\begin{subfigure}{0.48\linewidth}
    \centering
    \includegraphics[width=\linewidth]{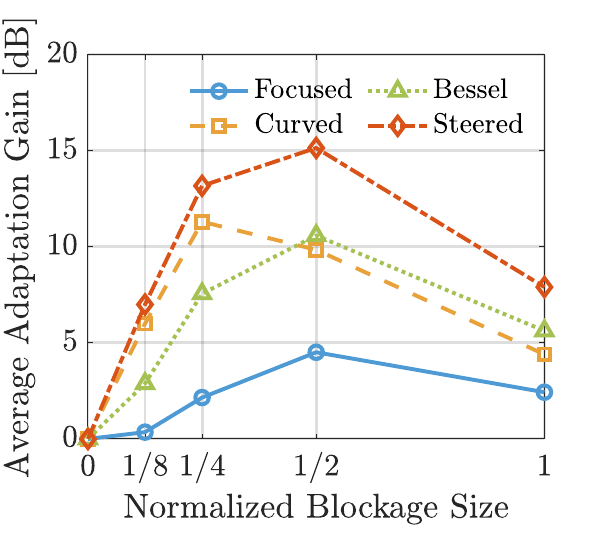}
    \caption{}
    \label{fig:avg-adaptation-gain-blockage-size}
\end{subfigure}

\caption{Event-level performance comparison for different blockage sizes, with blockage width normalized to the transmit aperture size $D$. (a) Adaptive upper-bound SNR. (b) Adaptation gain.}
\label{fig:event-level-blockage-size}
\end{figure}

Fig.~\ref{fig:avg-snr-blockage-size} shows the adaptive upper-bound SNR $\overline{\gamma}_s^{\mathrm{UB}}$ as the normalized blockage size increases. Observe that for all beam types, SNR decreases monotonically with increasing blockage size, indicating that even ideal beam-parameter adaptation is less effective under stronger obstruction. Nonetheless, the best beam type depends on blockage size. For small blockages, the curved beam achieves the highest average SNR, suggesting that trajectory control is effective when the blockage only partially intersects the main propagation region. For larger blockages, the focused beam becomes the strongest performer, outperforming the curved beam by approximately $2.5$ dB at $W_b/D=1/2$ and about $3$ dB at $W_b/D=1$. This suggests that when the obstruction is too large to route around, diffraction-assisted receiver coupling becomes more beneficial. Moreover, the Bessel beam outperforms the steered beam under smaller blockages due to partial self-healing, but degrades more rapidly as the blockage size grows and, perhaps surprisingly, eventually performs worse than the steered beam.

Fig.~\ref{fig:avg-adaptation-gain-blockage-size} shows the corresponding $G_s^{\mathrm{adapt}}$. The gain is zero when $W_b/D=0$ (no blockage) and for all beam types, the gain first increases as blockage becomes more severe, reaching its largest value at $W_b/D=1/2$ for all beam types except the curved beam, which peaks for a smaller blockage size. At blockage size $1/2$, the steered beam achieves the largest gain of about $13$ dB, followed by the Bessel and curved beams at roughly $10$ dB, while the focused beam has a smaller gain of about $4$ dB.
The focused beam maintains strong performance under larger blockages but gains less from adaptation because its fixed configuration is already resilient, whereas the steered beam achieves a large gain from a weaker baseline. The dominant mitigation mechanism also changes with blockage size: curved-beam trajectory control is most effective for smaller obstacles, whereas focused beams 
become more effective as the blockage size grows. Although the Bessel beam benefits from partial self-healing, this advantage diminishes under larger blockages. 

\textit{Findings: Significant adaptation gains are available during a blockage event as adaptation enables a transmitter to steer away from a receiver to realize diffraction gains. Transmitters can likewise benefit from adapting beam-specific parameters: the focused beam adapts by shortening the focal length, the Bessel beam increases its deflection angle and hence transverse energy distribution to improve self-healing, and the curved beam modifies its trajectory to bypass the obstruction. Nonetheless, while high adaptation gain seems favorable, comparing Figures \ref{fig:avg-snr-blockage-size} and \ref{fig:avg-adaptation-gain-blockage-size} shows that high adaptation gain does not correspond to the best average SNR. Moreover, realizing adaptation gains in a real system requires protocol support for measurement and feedback to guide re-configuration.}

\textbf{Effect of Blockage Distance:} We next study how blockage position relative to the transmitter and receiver affects event-level performance, as the blockage location determines how much distance remains for the beam to diffract, self-heal, or re-couple to the receiver.

\begin{figure}[t]
\centering

\begin{subfigure}{0.48\linewidth}
    \centering
    \includegraphics[width=\linewidth]{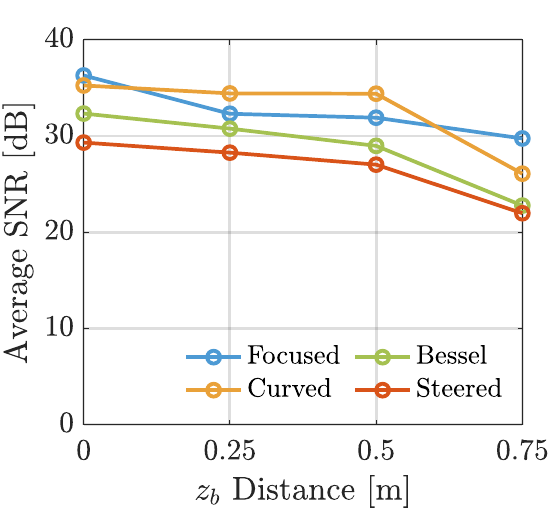}
    \caption{}
    \label{fig:avg-snr-z-distance}
\end{subfigure}
\hfill
\begin{subfigure}{0.48\linewidth}
    \centering
    \includegraphics[width=\linewidth]{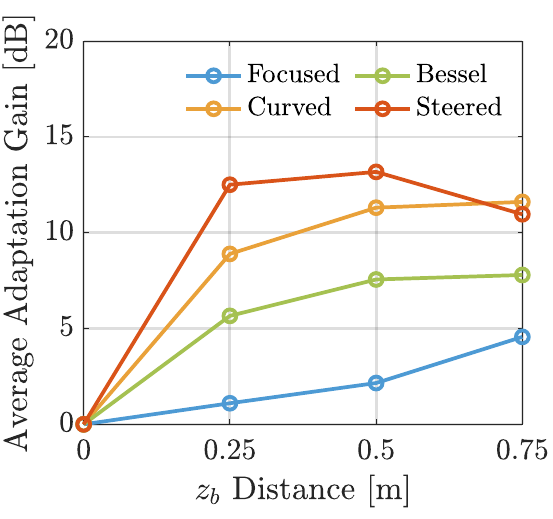}
    \caption{}
    \label{fig:avg-adaptation-gain-z-distance}
\end{subfigure}

\caption{Event-level performance comparison under different blockage locations along the propagation direction. (a) The average adaptive upper-bound SNR. (b) The average adaptation gain.}
\label{fig:event-level-z-distance}
\end{figure}

The receiver is again located at $1$~m from the transmitter, and the blockage is placed at different distances $z_b$ between the transmitter and receiver. We consider $z_b\in\{0,0.25,0.5,0.75\}$ m, where $z_b=0$ denotes the clear-channel case with no blockage. The blockage width is fixed, and the event-level metrics are computed over the blockage zone using $\tau=0.1$~dB. When $z_b=0$, no blockage is present. Thus, $\overline{\gamma}_s^{\mathrm{UB}}$ corresponds to the clear-channel SNR and $G_s^{\mathrm{adapt}}=0$.

Fig.~\ref{fig:avg-snr-z-distance} shows the adaptive upper-bound SNR $\overline{\gamma}_s^{\mathrm{UB}}$ as the blockage position is varied from the transmitter toward the receiver. The average SNR generally decreases as $z_b$ increases, indicating that blockage becomes more harmful when it occurs closer to the receiver. This is because less propagation distance remains for the field to diffract, self-heal, or re-couple energy back to the receiver.

The focused beam achieves high average SNR across most blockage locations and remains relatively stable for $z_b=0.25$~m and $z_b=0.5$~m. Its performance only drops more noticeably when the blockage is closer to the receiver at $z_b=0.75$~m, suggesting strong robustness from receiver-side energy concentration and diffraction-assisted coupling. The curved beam performs similarly to the focused beam at smaller and intermediate blockage distances, but degrades sharply at $z_b=0.75$~m, where there is limited distance for trajectory re-routing. The Bessel beam generally outperforms the steered beam due to partial self-healing, while the steered beam gives the lowest average SNR under blockage.

Fig.~\ref{fig:avg-adaptation-gain-z-distance} shows the corresponding $G_s^{\mathrm{adapt}}$. The adaptation gain is zero in the clear-channel case and increases once blockage is introduced. The steered beam achieves the largest gain, about $12$--$13$~dB, showing that steering-angle adaptation can recover a large amount of SNR from a weak fixed-beam baseline. The curved and Bessel beams also benefit from adaptation, with gains of roughly $10$ dB and $7$ dB, respectively. In contrast, the focused beam has the smallest gain, increasing from about $1$~dB at $z_b=0.25$~m to about $4$--$5$~dB at $z_b=0.75$~m.

\textit{Findings: This analysis, again, shows that the beam with the largest adaptation gain does not necessarily achieve the highest adapted SNR. The focused beam maintains the strongest performance across blockage locations and becomes increasingly advantageous near the receiver, where limited propagation distance reduces the effectiveness of trajectory re-routing and self-healing. In contrast, the steered beam achieves a large gain from a weak fixed-beam baseline, while the curved and Bessel beams rely more strongly on sufficient distance for trajectory adjustment and field reconstruction. Thus, practical beam selection should consider blockage location, final blocked-link SNR, adaptation gain, and the associated adaptation overhead.}

\subsection{Adaptive Beam-Parameter Selection}

\begin{figure}[t]
\centering
\includegraphics[width=1\linewidth]{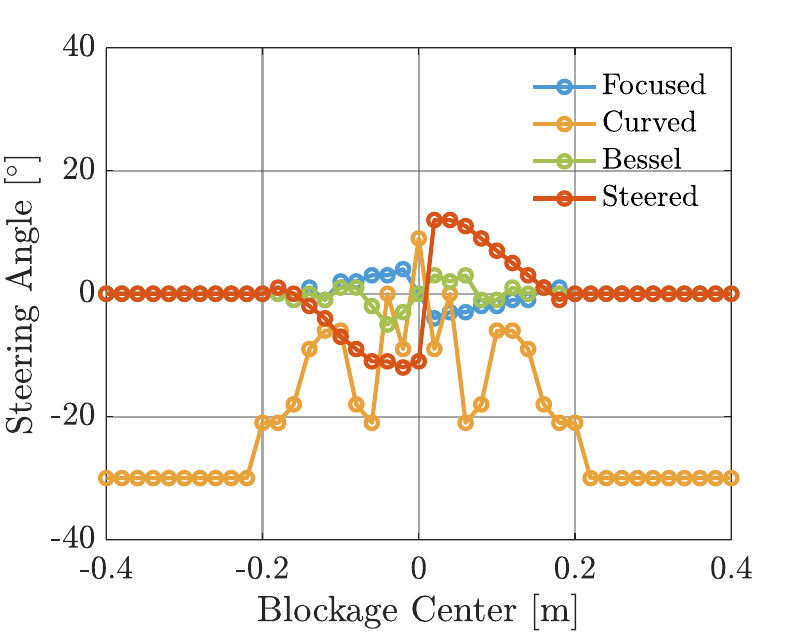}
\caption{Steering angle of different near-field beam types under the adaptive upper-bound policy. The blockage is placed at $z_b=0.5$~m, with normalized blockage width $W_b/D=1/4$.}
\label{fig:beam-shape-comparison-theta0}
\end{figure}

Finally, to study how different beam types adapt their parameters to circumvent blockage, we examine the parameters selected by the adaptive upper-bound policy throughout a blockage event. The blockage is located at $z_b=0.5$~m, with normalized width $W_b/D=1/4$, and the receiver is located at $z_u=1$~m. All other settings follow the previous subsection. At each blockage center position, the policy selects the beam-parameter combination that maximizes the received SNR.

Fig.~\ref{fig:beam-shape-comparison-theta0} shows the selected steering angle for each beam type. Outside the effective blockage zone, the angles remain close to their clear-channel optima. As the blockage intersects the dominant propagation region, the policy adjusts the steering angles to redirect energy around the obstruction and improve receiver coupling. The steered beam exhibits the largest angular variation because steering is its only degree of freedom. In contrast, the focused and Bessel beams require smaller steering adjustments because they can also adapt their beam-specific parameters. The curved beam exhibits a less regular but approximately symmetric pattern, reflecting the joint optimization of its steering angle and trajectory parameters.

\begin{figure}[t]
\centering

\begin{subfigure}{0.48\linewidth}
    \centering
    \includegraphics[width=\linewidth]{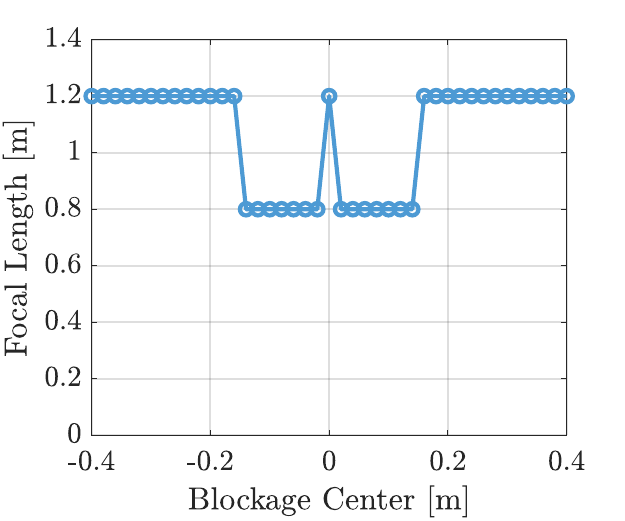}
    \caption{Focused: focal length.}
    \label{fig:focused-focal-length-adaptive}
\end{subfigure}
\hfill
\begin{subfigure}{0.48\linewidth}
    \centering
    \includegraphics[width=\linewidth]{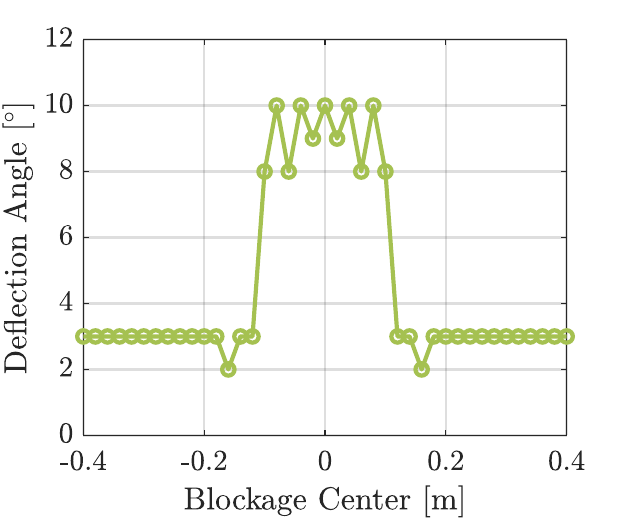}
    \caption{Bessel: deflection angle.}
    \label{fig:bessel-deflection-angle-adaptive}
\end{subfigure}

\vspace{0.8em}

\begin{subfigure}{1\linewidth}
    \centering
    \includegraphics[width=\linewidth]{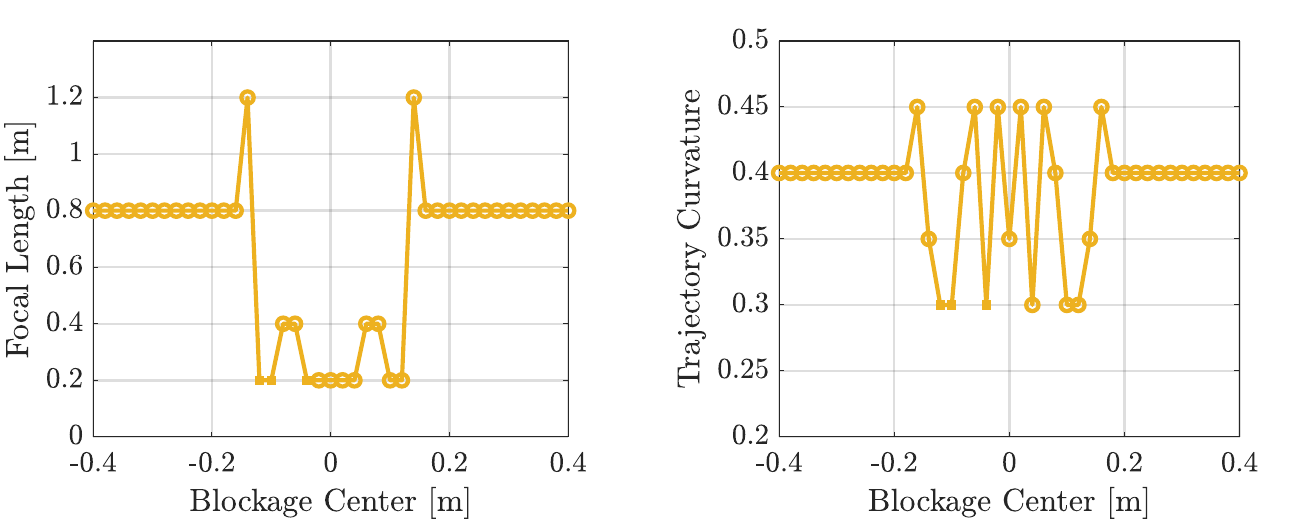}
    \caption{Curved: focal length and curvature parameter.}
    \label{fig:curved-parameters-adaptive}
\end{subfigure}

\caption{Adaptive beam parameters selected by the adaptive upper-bound policy during a blockage event.}
\label{fig:adaptive-parameter-selection}
\end{figure}

Fig.~\ref{fig:adaptive-parameter-selection} shows how the beam-specific parameters change throughout the blockage event. Outside the effective blockage zone, the parameters remain at their clear-channel values. For the focused beam, Fig.~\ref{fig:focused-focal-length-adaptive} shows that the focal length decreases from $1.2$~m to $0.8$~m for most off-center blockage positions, moving the focal region upstream to improve diffraction-assisted receiver coupling. It briefly returns to $1.2$~m at $x_b=0$, indicating that shortening the focal length is primarily beneficial under asymmetric blockage. For the Bessel beam, Fig.~\ref{fig:bessel-deflection-angle-adaptive} shows that the deflection angle increases from $3^\circ$ to approximately $8^\circ$--$10^\circ$ when the blockage overlaps the beam core, redistributing energy transversely to promote self-healing. For the curved beam, Fig.~\ref{fig:curved-parameters-adaptive} shows that the focal length varies between approximately $0.2$ and $1.2$~m, while the curvature parameter varies between $0.3$ and $0.45$. These joint adjustments produce position-dependent trajectories that re-route energy around the blockage. The symmetric but nonmonotonic patterns reflect switching among the discrete parameter combinations evaluated by the adaptive upper-bound search.

\textit{Findings: The adaptation gains arise from distinct beam-specific mechanisms. The focused beam shortens its focal length to improve diffraction-assisted receiver coupling, the Bessel beam increases its deflection angle to promote self-healing, and the curved beam jointly adjusts its focal length and curvature to re-route its trajectory. In contrast, the steered beam relies entirely on angular redirection, explaining its larger steering variations and substantial gain from a weak fixed-beam baseline. Thus, effective blockage-aware adaptation should exploit beam-specific degrees of freedom rather than relying solely on directional steering.}

\section{Experimental Evaluation}
In this section, we experimentally evaluate the candidate beam types under dynamic blockage. We first introduce the measurement setup and metasurface-based beam realization, and then present the measured blockage response and adaptation results for the considered blockage scenarios.

\subsection{Sub-THz Measurement Testbed}
\begin{figure}[t]
    \centering
    \includegraphics[width=0.85\linewidth]{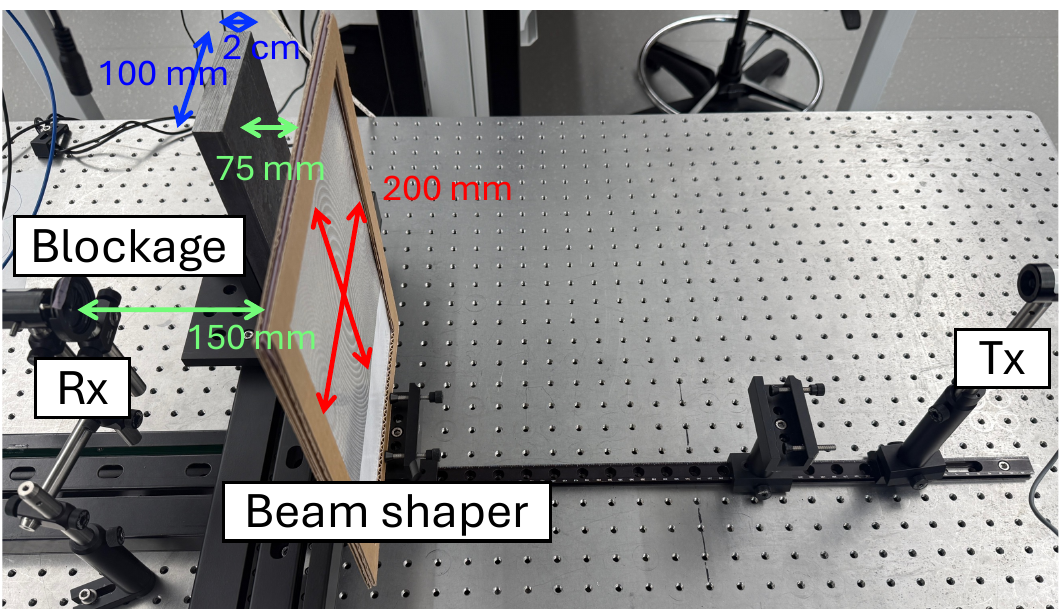}
    \caption{Sub-THz experimental testbed for near-field blockage measurements.}
    \label{fig:experimental-testbed}
\end{figure}

We characterize the blockage response of the considered beam types using a Toptica TeraFlash terahertz time-domain spectroscopy system. The platform consists of fiber-coupled transmitter and receiver heads that generate broadband THz pulses and measure the received response as illustrated in Fig.~\ref{fig:experimental-testbed}. The measured time-domain signal is transformed into the frequency domain, from which we extract the received power at the $150$~GHz design frequency. This measurement procedure enables a controlled comparison of how each beam type responds to the same moving blockage and how the adapted beam configurations recover the blocked link.

In the setup, the transmitter is placed $450$ mm from the beam shaper to illuminate the full $200$ mm $\times$ $200$ mm aperture. The beam shaper is fixed normal to the incident field, and the receiver is positioned at broadside, $150$ mm from the beam shaper, corresponding to the target receive location. To emulate a dynamic blockage event, a $100$ mm $\times$ $20$ mm metal blockage is mounted on a linear translation stage, positioned $75$ mm from the beam shaper, and translated laterally across the beam path with controlled blockage positions. This small spatial scale is due to the submicrowatt power of the transmitter, which severely constrains the link budget. 

\subsection{Metasurface-Based Beam Realization}
While future sub-THz systems may use dense, electronically reconfigurable arrays or programmable metasurfaces to generate near-field beam profiles, such hardware is still in an early stage of development \cite{venkatesh2020high}. We therefore use passive transmissive metasurface beam shapers as a practical experimental platform for realizing the desired aperture phase profiles. In particular, we use C-shaped split-ring resonators (CSRRs) as meta-atoms because they provide controllable phase responses in the sub-THz band while maintaining a relatively stable transmission amplitude \cite{zhang2013broadband,shaikhanov2022metasurface}. The metasurface is designed for a center frequency of $150$ GHz and enables us to realize the beam-shaping phase profiles defined in Section III.

\begin{figure}
    \centering

    \begin{subfigure}{0.46\linewidth}
        \centering
        \includegraphics[width=\linewidth]{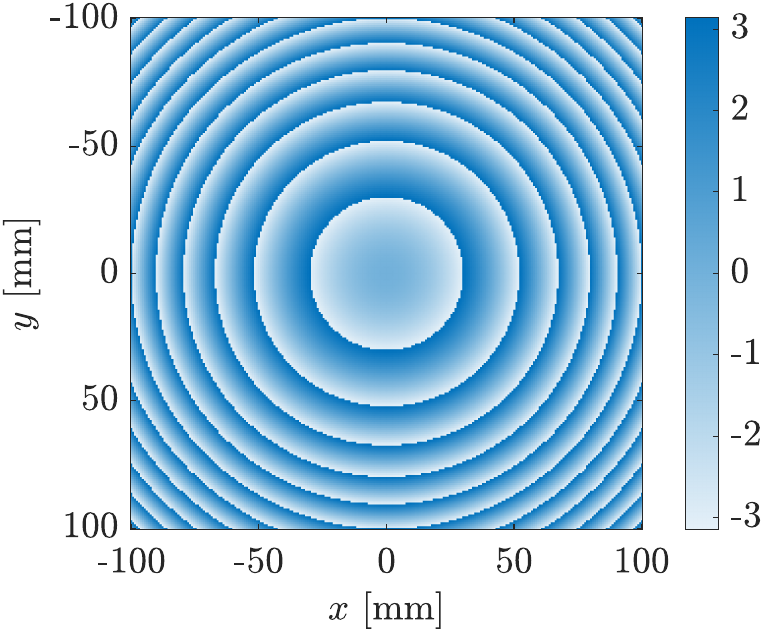}
        \caption{Incident phase}
        \label{fig:phase-compensation-incident}
    \end{subfigure}
    \hfill
    \begin{subfigure}{0.46\linewidth}
        \centering
        \includegraphics[width=\linewidth]{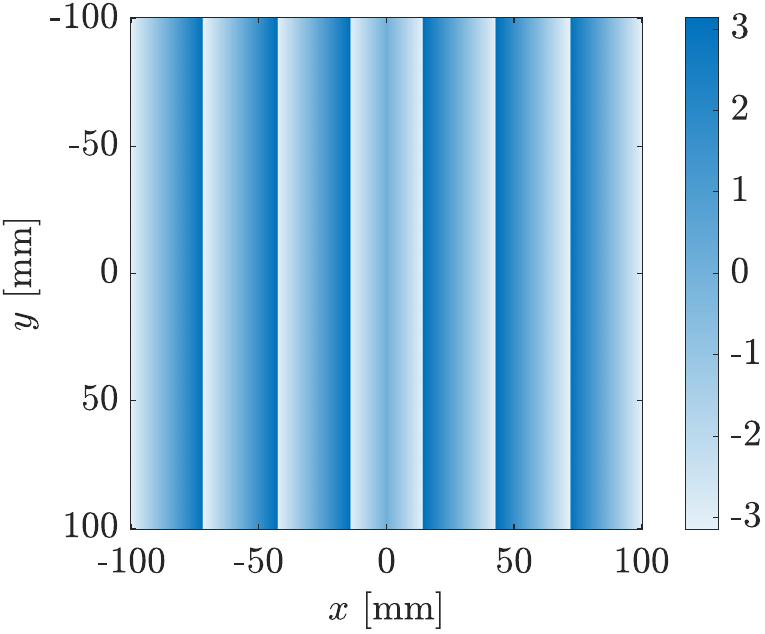}
        \caption{Desired outgoing phase}
        \label{fig:phase-compensation-desired}
    \end{subfigure}

    \vspace{0.5em}

    \begin{subfigure}{0.46\linewidth}
        \centering
        \includegraphics[width=\linewidth]{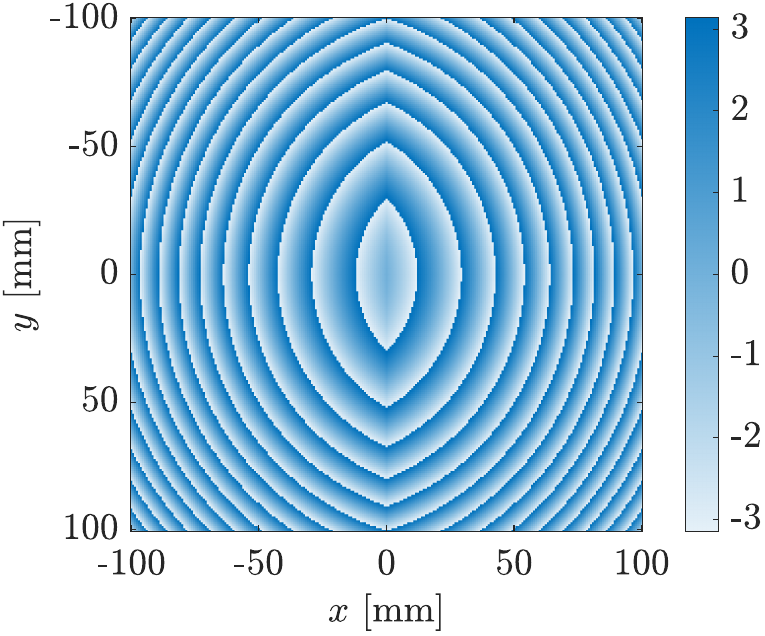}
        \caption{Compensated phase}
        \label{fig:phase-compensation-ms}
    \end{subfigure}
    \hfill
    \begin{subfigure}{0.46\linewidth}
        \centering
        \includegraphics[width=\linewidth]{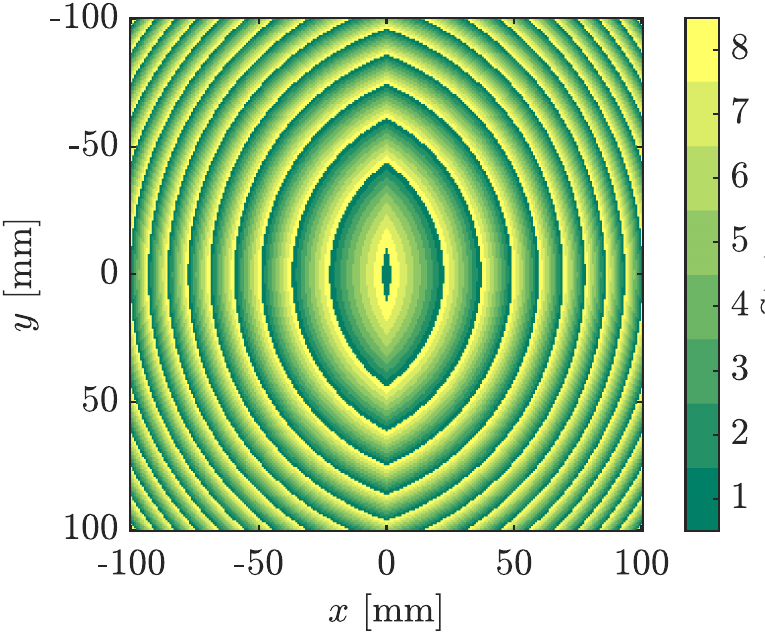}
        \caption{Transmitted state}
        \label{fig:phase-compensation-transmitted}
    \end{subfigure}

    \caption{Example of phase pre-compensation for a Bessel beam at the beam-shaper plane.}
    \label{fig:phase-compensation}
\end{figure}

Because the transmitter is located in the near-field of the beam shaper, the incident field at the metasurface is not perfectly planar. We therefore pre-compensate the implemented phase profile to cancel the incident spherical wavefront at the beam-shaper plane, as illustrated in Fig.~\ref{fig:phase-compensation}. Let $\phi_{n,s}(\Theta_s)$ denote the desired outgoing phase profile for beam type $s$ at the $n$-th aperture location, and let $\phi_n^{\mathrm{comp}}$ denote the spherical-wave compensation term. The continuous phase implemented by the metasurface is
\begin{equation}
    \phi_{n,s}^{\mathrm{MS}}(\Theta_s)
    =
    \left[
    \phi_{n,s}(\Theta_s)
    +
    \phi_n^{\mathrm{comp}}
    \right]_{2\pi},
\end{equation}
where $[\cdot]_{2\pi}$ denotes phase wrapping modulo $2\pi$. The compensation term is computed from the path-length difference between the transmitter and each beam-shaper location. If the transmitter is centered on the beam-shaper axis and separated from the beam shaper by a distance $L$, then
\begin{equation}
    \phi_n^{\mathrm{comp}}
    =
    -k
    \left(
    \sqrt{x_n^2+L^2}-L
    \right).
\end{equation}

This compensation ensures that, after the incident spherical wavefront interacts with the metasurface, the transmitted field follows the intended beam-shaping phase profile.

After pre-compensation, the continuous metasurface phase is quantized into the available discrete CSRR phase states. In our implementation, the phase profile is quantized into eight states, with an approximately $45^\circ$ phase difference between adjacent states to cover the full $2\pi$ phase range while maintaining near-constant transmission amplitude. Each state is realized using a distinct CSRR geometry~\cite{liao2025terafocus}. The metasurface patterns are then fabricated using a hot-stamping process, which provides a low-cost and rapid method for prototyping metallic sub-THz metasurfaces~\cite{guerboukha2021high,liao2025terafocus}.

\subsection{Experimental Results}
Here, we experimentally study beam types and adaptation under blockage. First, we experimentally characterize the beam-type-dependent blockage loss induced by the same physical blockage. Second, we quantify how much lost link performance can be recovered when the beam configuration is changed according to the adaptive policy.

\textit{Scenarios:} For each beam type, we evaluate three cases. The first case is the unblocked fixed-beam case, which measures the received SNR without the blockage and serves as the clear-channel reference. In this case, each beam uses its fixed clear-channel configuration as optimized by the ASM-based parameter sweeping method described in Section III: the focused beam has $l=150$~mm and $\theta=0^\circ$, the curved beam has $\theta=-20^\circ$, $l=120$~mm, and $\kappa=0.5$, the Bessel beam has $\psi_b=2^\circ$ and $\theta=0^\circ$, and the steered beam has $\theta=0^\circ$. The second scenario introduces a blockage and uses the \emph{adaptation-free policy}. In particular, the same clear-channel-optimized beam configurations are used after the metal blockage is inserted. The blockage position is selected based on the fixed curved beam's largest SNR degradation, creating a representative challenging blockage condition for comparing blockage loss across beam types without adaptation. The third scenario evaluates the \emph{adaptive upper-bound policy}. Here, the adaptive beam parameters are selected numerically for the blockage condition via the method described in Section III, i.e., ASM-based exhaustive search to find the parameters that maximize SNR. We then experimentally realize the adaptive beams by fabricating metasurfaces that correspond to the required phase profiles. For the selected blockage position, the optimized adaptive parameters for each beam type are as follows: the focused beam has $l=80$~mm and $\theta=3^\circ$, the curved beam has $\theta=0^\circ$, $l=50$~mm, and $\kappa=0.1$, the Bessel beam has $\psi_b=5^\circ$ and $\theta=-2^\circ$, and the steered beam has $\theta=-10^\circ$. Thus, since the fabricated metasurfaces are passive, adaptation is emulated by fabricating different metasurface configurations rather than by real-time electronic reconfiguration.

\begin{figure}
\centering

\begin{subfigure}{1\linewidth}
    \centering
    \includegraphics[width=\linewidth]{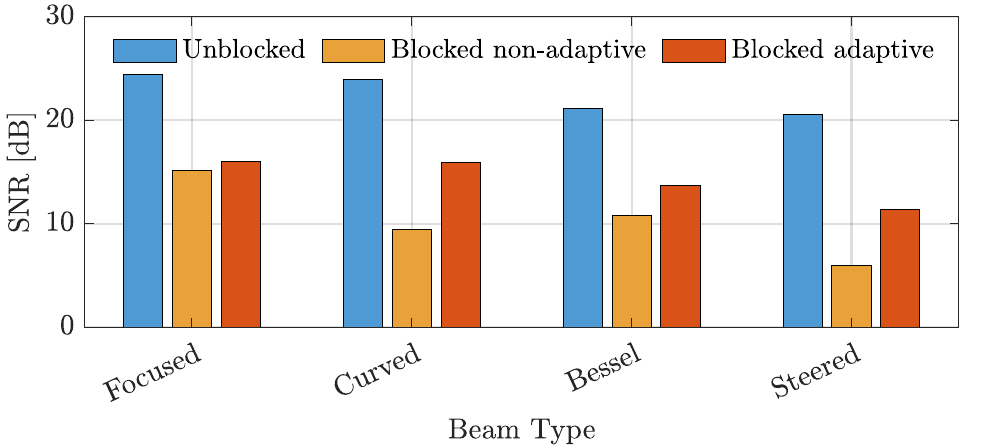}
    \caption{Experimental results}
    \label{fig:experimental-results}
\end{subfigure}

\vspace{0.6em}

\begin{subfigure}{1\linewidth}
    \centering
    \includegraphics[width=\linewidth]{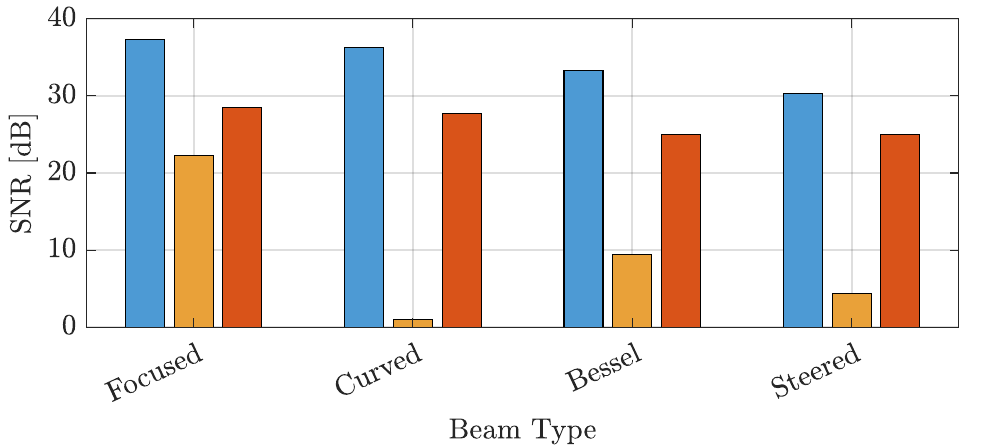}
    \caption{Simulated results}
    \label{fig:simulation-results}
\end{subfigure}

\caption{Experimental and simulated blockage results under the considered measurement scenario.}
\label{fig:experiment-simulation-results}

\end{figure}

Fig.~\ref{fig:experimental-results} shows the experimentally measured SNR for the four beam types in each of the three scenarios above: the unblocked case, the blocked case without adaptation (adaptation-free policy) and the blocked case with adaptation (adaptive upper-bound policy). In the unblocked case (blue bars), the focused and curved beams achieve the highest clear-channel SNRs at approximately $25$~dB, while the Bessel and steered beams start from lower baselines near $20$~dB. After the metal blockage is introduced, the SNR decreases for all beam types. When the beams do not adapt to the obstacle (orange bars), i.e., the non-adaptive policy, the focused beam experiences the smallest degradation, decreasing to approximately $15$~dB, followed by the Bessel beam at $10.5$~dB. The curved and steered beams decrease to approximately $9.5$~dB and $6$~dB, respectively. Finally, with the adaptive upper-bound policy, all beams partially recover losses due to the obstacle. For example, the focused beam has an adaptation gain of approximately $1$~dB, while the curved, Bessel, and steered beams have adaptation gains of $6.5$~dB, $3$~dB, and $5$~dB, respectively. Note that with adaptation, the focused and curved beams achieve nearly identical performance ($16.0$~dB and $15.9$~dB, respectively). Unfortunately, even with adaptation, the Bessel beam cannot achieve performance comparable to that of the focused and curved beams, as its link is limited to $13.7$~dB SNR. Nonetheless, the adapted Bessel beam still outperforms the adapted steered beam, which achieves only 11.5 dB.

\textit{Findings: For the selected blockage condition, the focused beam provides the best performance without adaptation. Indeed, when beam adaptation to obstacles is not possible, focused beams surprisingly perform best in the cases studied, despite lacking obstacle-resilience properties (e.g., self-healing). In contrast, when each beam's parameters are re-optimized according to the obstacle, all beam types improve performance. In this case, the curved beam is able to reconfigure its trajectory to largely avoid the obstacle. However, the focused beam again performs surprisingly well, with an SNR nearly identical to that of the adapted curved beam ($16.0$~dB and $15.9$~dB, respectively). Perhaps equally surprising is that they both significantly outperform the adapted Bessel beam (which achieves $13.7$~dB SNR), indicating that the self-healing property alone does not guarantee the highest blocked-link SNR.}

Fig.~\ref{fig:simulation-results} shows the corresponding simulation results for the same geometry. The simulation exhibits the same qualitative ordering as the measurements but with a larger dynamic range. The focused beam again shows the smallest fixed-beam degradation, while the curved beam experiences the largest fixed-beam loss and the greatest recovery after adaptation. The Bessel and steered beams also recover through adaptation, but their final SNRs remain below those of the focused and curved beams. Compared with the experiment, the simulated clear-channel SNRs and adaptation gains are larger.

\textit{Findings:
For the considered geometry, similarities between simulation and experimental results indicate that the framework captures the main beam-dependent blockage trends, including surprising resilience of focused beams, curved-beam sensitivity to trajectory interception, and beam-specific recovery through adaptation. The lower experimental SNR and adaptation gains are attributable to practical implementation losses, such as metasurface phase quantization, fabrication tolerances, finite-aperture effects, alignment errors, and measurement nonidealities. Therefore, the simulation provides a reliable basis for comparing the relative behavior of different beam types, although it may overestimate the absolute performance achievable with practical hardware.}

\section{Conclusion}

This paper presented an empirical analysis of near-field beam shaping under dynamic blockage events. By evaluating representative beam types and their corresponding parameters under two transmission policies, the framework enables a comparison of structured beam behaviors throughout various blockage events. The results show that blockage resilience depends not on a single beam type or its parameters, but strongly on beam characteristics, blockage geometry, and beam adaptation. Consequently, future blockage-aware communication protocols should jointly optimize beam type selection and beam parameter adaptation to the operating environment, balancing the associated sensing, channel estimation, feedback, optimization, and hardware reconfiguration overhead. We envision the proposed analysis as serving as a benchmark for the design and evaluation of practical blockage-aware near-field beam adaptation strategies.

\bibliographystyle{IEEEtran}
\bibliography{references}

@article{akyildiz2014terahertz,
  author  = {I. F. Akyildiz and J. M. Jornet and C. Han},
  title   = {Terahertz band: Next frontier for wireless communications},
  journal = {Physical Communication},
  volume  = {12},
  pages   = {16--32},
  year    = {2014},
  doi     = {10.1016/j.phycom.2014.01.006}
}

@article{koenig2013wireless,
  author  = {S. Koenig and D. Lopez-Diaz and J. Antes and F. Boes
             and R. Henneberger and A. Leuther and A. Tessmann
             and R. Schmogrow and D. Hillerkuss and R. Palmer
             and T. Zwick and C. Koos and W. Freude and O. Ambacher
             and J. Leuthold and I. Kallfass},
  title   = {Wireless sub-{THz} communication system with high data rate},
  journal = {Nature Photonics},
  volume  = {7},
  number  = {12},
  pages   = {977--981},
  year    = {2013},
  doi     = {10.1038/nphoton.2013.275}
}

@article{shurakov2023empirical,
  author  = {A. Shurakov and D. Moltchanov and A. Prikhodko
             and A. Khakimov and E. Mokrov and V. Begishev
             and I. Belikov and Y. Koucheryavy and G. Gol'tsman},
  title   = {Empirical blockage characterization and detection in indoor
             sub-{THz} communications},
  journal = {Computer Communications},
  volume  = {201},
  pages   = {48--58},
  year    = {2023},
  doi     = {10.1016/j.comcom.2023.01.017}
}

@article{shurakov2023dynamic,
  author  = {A. Shurakov and P. Rozhkova and A. Khakimov
             and E. Mokrov and A. Prikhodko and V. Begishev
             and Y. Koucheryavy and M. Komarov and G. Gol'tsman},
  title   = {Dynamic blockage in indoor reflection-aided sub-terahertz
             wireless communications},
  journal = {IEEE Access},
  volume  = {11},
  pages   = {134677--134689},
  year    = {2023},
  doi     = {10.1109/ACCESS.2023.3337050}
}

@article{doeker2025human,
  author  = {T. Doeker and M. Eggers and C. E. Reinhardt
             and D. M. Mittleman and T. K{\"u}rner},
  title   = {Human motion sensing through blockage and reflection
             measurements at 60 {GHz} and 300 {GHz}},
  journal = {IEEE Access},
  volume  = {13},
  pages   = {97997--98005},
  year    = {2025},
  doi     = {10.1109/ACCESS.2025.3573681}
}

@article{petrov2023mobile,
  author  = {V. Petrov and D. Bodet and A. Singh},
  title   = {Mobile near-field terahertz communications for {6G} and
             {7G} networks: Research challenges},
  journal = {Frontiers in Communications and Networks},
  volume  = {4},
  pages   = {1151324},
  year    = {2023},
  doi     = {10.3389/frcmn.2023.1151324}
}

@article{singh2023wavefront,
  author  = {Singh, Arjun and Petrov, Vitaly and Guerboukha, Hichem
          and Reddy, Innem V. A. K. and Knightly, Edward W.
          and Mittleman, Daniel M. and Jornet, Josep M.},
  title   = {Wavefront engineering: Realizing efficient terahertz band
             communications in {6G} and beyond},
  journal = {IEEE Wireless Communications},
  volume  = {31},
  number  = {3},
  pages   = {133--139},
  year    = {2024},
  doi     = {10.1109/MWC.019.2200583}
}

@article{zhang20236g,
  author  = {H. Zhang and N. Shlezinger and F. Guidi and D. Dardari
             and Y. C. Eldar},
  title   = {{6G} wireless communications: From far-field beam steering
             to near-field beam focusing},
  journal = {IEEE Communications Magazine},
  volume  = {61},
  number  = {4},
  pages   = {72--77},
  year    = {2023},
  doi     = {10.1109/MCOM.001.2200259}
}

@article{nepa2017near,
  author  = {P. Nepa and A. Buffi},
  title   = {Near-field-focused microwave antennas: Near-field shaping
             and implementation},
  journal = {IEEE Antennas and Propagation Magazine},
  volume  = {59},
  number  = {3},
  pages   = {42--53},
  year    = {2017},
  doi     = {10.1109/MAP.2017.2686118}
}

@inproceedings{liao2025terafocus,
  author    = {J.-C. Liao and B. Bilgin and E. W. Knightly},
  title     = {{TeraFocus}: Wideband beam focusing with radial mobility},
  booktitle = {Proc. 26th ACM Int. Symp. Theory, Algorithmic Foundations,
               and Protocol Design for Mobile Networks and Mobile
               Computing (MobiHoc)},
  pages     = {81--90},
  year      = {2025},
  doi       = {10.1145/3704413.3764418}
}

@article{durnin1988comparison,
  author  = {J. Durnin and Miceli, Jr., J. J. and J. H. Eberly},
  title   = {Comparison of {Bessel} and {Gaussian} beams},
  journal = {Optics Letters},
  volume  = {13},
  number  = {2},
  pages   = {79--80},
  year    = {1988},
  doi     = {10.1364/OL.13.000079}
}

@article{reddy2023ultrabroadband,
  author  = {Reddy, Innem V. A. K. and D. Bodet and A. Singh and V. Petrov
             and C. Liberale and J. M. Jornet},
  title   = {Ultrabroadband terahertz-band communications with
             self-healing {Bessel} beams},
  journal = {Communications Engineering},
  volume  = {2},
  number  = {1},
  pages   = {70},
  year    = {2023},
  doi     = {10.1038/s44172-023-00118-8}
}

@article{bodet2024sub,
  author  = {D. Bodet and V. Petrov and S. Petrushkevich and J. M. Jornet},
  title   = {Sub-terahertz near field channel measurements and analysis
             with beamforming and {Bessel} beams},
  journal = {Scientific Reports},
  volume  = {14},
  number  = {1},
  pages   = {19675},
  year    = {2024},
  doi     = {10.1038/s41598-024-70542-z}
}

@article{guerboukha2024curving,
  author  = {H. Guerboukha and B. Zhao and Z. Fang and E. W. Knightly
             and D. M. Mittleman},
  title   = {Curving {THz} wireless data links around obstacles},
  journal = {Communications Engineering},
  volume  = {3},
  number  = {1},
  pages   = {58},
  year    = {2024},
  doi     = {10.1038/s44172-024-00206-3}
}

@article{wang2026blockage,
  author  = {Y. Wang and L. Dai},
  title   = {Blockage-robust beamforming for near-field communications:
             From single-{Airy} to multi-{Airy}},
  journal = {arXiv preprint arXiv:2607.07278},
  year    = {2026}
}

@article{chen2025physics,
  author  = {H. Chen and A. Kludze and Y. Ghasempour},
  title   = {A physics-informed {Airy} beam learning framework for
             blockage avoidance in sub-terahertz wireless networks},
  journal = {Nature Communications},
  volume  = {16},
  number  = {1},
  pages   = {7387},
  year    = {2025},
  doi     = {10.1038/s41467-025-62443-0}
}

@article{uchimura2025optimization,
  author  = {S. Uchimura and J. M. Jornet and K. Ishibashi},
  title   = {Optimization and characterization of near-field beams
             with uniform linear arrays},
  journal = {IEEE Transactions on Wireless Communications},
  volume  = {25},
  pages   = {6904--6920},
  year    = {2026},
  doi     = {10.1109/TWC.2025.3627599}
}

@article{uchimura2026optimal,
  author  = {S. Uchimura and J. M. Jornet and K. Ishibashi},
  title   = {Optimal wavefronts for maximum ratio transmissions under
             path blockage effects},
  journal = {IEEE Transactions on Wireless Communications},
  volume  = {25},
  pages   = {11051--11067},
  year    = {2026},
  doi     = {10.1109/TWC.2026.3658566}
}

@article{weng2025learning,
  author  = {C. Weng and Y. Guo and B. Zhao and Y. Wang and W. Chen
             and Z. Li},
  title   = {Learning-based blockage-resilient beam training in
             near-field terahertz communications},
  journal = {IEEE Transactions on Wireless Communications},
  volume  = {25},
  pages   = {20945--20961},
  year    = {2026},
  doi     = {10.1109/TWC.2026.3716726}
}

@article{zhang2026breaking,
  author  = {S. Zhang and B. Di and L. Song},
  title   = {Breaking near-field communication barriers: Focused,
             curved, or {Airy} beamforming?},
  journal = {arXiv preprint arXiv:2604.01704},
  year    = {2026}
}

@article{arora2022efficient,
  author  = {A. Arora and C. G. Tsinos and M. R. Bhavani Shankar
             and S. Chatzinotas and B. Ottersten},
  title   = {Efficient algorithms for constant-modulus analog beamforming},
  journal = {IEEE Transactions on Signal Processing},
  volume  = {70},
  pages   = {756--771},
  year    = {2022},
  doi     = {10.1109/TSP.2021.3094653}
}

@article{prado2022nearfield,
  author  = {D. R. Prado},
  title   = {Near field models of spatially-fed planar arrays and their
             application to multi-frequency direct layout optimization
             for mm-wave {5G New Radio} indoor network coverage},
  journal = {Sensors},
  volume  = {22},
  number  = {22},
  pages   = {8925},
  year    = {2022},
  doi     = {10.3390/s22228925}
}

@article{wu2023multiple,
  author  = {Z. Wu and L. Dai},
  title   = {Multiple access for near-field communications:
             {SDMA} or {LDMA}?},
  journal = {IEEE Journal on Selected Areas in Communications},
  volume  = {41},
  number  = {6},
  pages   = {1918--1935},
  year    = {2023},
  doi     = {10.1109/JSAC.2023.3275616}
}

@inproceedings{hassan2026multi,
  author    = {F. Hassan and J.-C. Liao and S. N. Jovanovi{\'c}
               and E. W. Knightly},
  title     = {{Multi-Spotlight}: A system for sub-{THz} multi-user
               networking},
  booktitle = {Proc. IEEE INFOCOM 2026---IEEE Conf. Computer Communications},
  pages     = {1--10},
  year      = {2026},
  doi       = {10.1109/INFOCOM59046.2026.11571437}
}

@article{bouchal1998self,
  author  = {Z. Bouchal and J. Wagner and M. Chlup},
  title   = {Self-reconstruction of a distorted nondiffracting beam},
  journal = {Optics Communications},
  volume  = {151},
  number  = {4--6},
  pages   = {207--211},
  year    = {1998},
  doi     = {10.1016/S0030-4018(98)00085-6}
}

@article{aiello2014wave,
  author  = {A. Aiello and G. S. Agarwal},
  title   = {Wave-optics description of self-healing mechanism in
             {Bessel} beams},
  journal = {Optics Letters},
  volume  = {39},
  number  = {24},
  pages   = {6819--6822},
  year    = {2014},
  doi     = {10.1364/OL.39.006819}
}

@book{born2013principles,
  author    = {M. Born and E. Wolf},
  title     = {Principles of Optics: Electromagnetic Theory of Propagation,
               Interference and Diffraction of Light},
  edition   = {7th},
  address   = {Cambridge, U.K.},
  publisher = {Cambridge University Press},
  year      = {1999},
  doi       = {10.1017/CBO9781139644181}
}

@article{petrov2024wavefront,
  author  = {V. Petrov and H. Guerboukha and A. Singh and J. M. Jornet},
  title   = {Wavefront hopping for physical layer security in {6G} and
             beyond near-field {THz} communications},
  journal = {IEEE Transactions on Communications},
  volume  = {73},
  number  = {5},
  pages   = {2996--3012},
  year    = {2025},
  doi     = {10.1109/TCOMM.2024.3484937}
}

@inproceedings{yazdnian2025nirvawave,
  author    = {V. Yazdnian and Y. Ghasempour},
  title     = {{NirvaWave}: An accurate and efficient near field wave
               propagation simulator for {6G} and beyond},
  booktitle = {Proc. IEEE Wireless Communications and Networking
               Conference (WCNC)},
  pages     = {1--7},
  year      = {2025},
  doi       = {10.1109/WCNC61545.2025.10978761}
}

@article{venkatesh2020high,
  author  = {S. Venkatesh and X. Lu and H. Saeidi and K. Sengupta},
  title   = {A high-speed programmable and scalable terahertz holographic
             metasurface based on tiled {CMOS} chips},
  journal = {Nature Electronics},
  volume  = {3},
  number  = {12},
  pages   = {785--793},
  year    = {2020},
  doi     = {10.1038/s41928-020-00497-2}
}

@article{zhang2013broadband,
  author  = {X. Zhang and Z. Tian and W. Yue and J. Gu and S. Zhang
             and J. Han and W. Zhang},
  title   = {Broadband terahertz wave deflection based on {C}-shape
             complex metamaterials with phase discontinuities},
  journal = {Advanced Materials},
  volume  = {25},
  number  = {33},
  pages   = {4567--4572},
  year    = {2013},
  doi     = {10.1002/adma.201204850}
}

@inproceedings{shaikhanov2022metasurface,
  author    = {Z. Shaikhanov and F. Hassan and H. Guerboukha
               and D. M. Mittleman and E. W. Knightly},
  title     = {Metasurface-in-the-middle attack:
               From theory to experiment},
  booktitle = {Proc. 15th ACM Conf. Security and Privacy in Wireless
               and Mobile Networks (WiSec)},
  pages     = {257--267},
  year      = {2022},
  doi       = {10.1145/3507657.3528549}
}

@article{guerboukha2021high,
  author  = {Guerboukha, Hichem and Amarasinghe, Yasith and Shrestha, Rabi and Pizzuto, Angela and Mittleman, Daniel M.},
  title   = {High-volume rapid prototyping technique for terahertz metallic metasurfaces},
  journal = {Optics Express},
  volume  = {29},
  number  = {9},
  pages   = {13806--13814},
  year    = {2021},
  doi     = {10.1364/OE.422991}
}

@book{goodman2017introduction,
  author    = {Goodman, Joseph W.},
  title     = {Introduction to Fourier Optics},
  edition   = {4th},
  publisher = {W. H. Freeman and Company},
  address   = {New York, NY, USA},
  year      = {2017}
}

@inproceedings{maccartney2017rapid,
  author    = {MacCartney, Jr., George R. and Rappaport, Theodore S. and Rangan, Sundeep},
  title     = {Rapid fading due to human blockage in pedestrian crowds at {5G} millimeter-wave frequencies},
  booktitle = {2017 IEEE Global Communications Conference (GLOBECOM)},
  pages     = {1--7},
  year      = {2017},
  doi       = {10.1109/GLOCOM.2017.8254900}
}

@inproceedings{islam2025blockage,
  author    = {Islam, Md Hasibul and Petrov, Vitaly and Guerboukha, Hichem},
  title     = {Blockage mitigation via curved {Airy} beams in near field terahertz communications beyond {6G}},
  booktitle = {2025 IEEE 22nd Consumer Communications \& Networking Conference (CCNC)},
  pages     = {1--6},
  year      = {2025},
  doi       = {10.1109/CCNC54725.2025.10975973}
}





\end{document}